# Heavy particle radioactivity from superheavy nuclei leading to $^{298}$114 daughter nuclei

K. P. Santhosh* and B. Priyanka
*School of Pure and Applied Physics, Kannur University, Swami Anandatheertha Campus, Payyanur 670 327, Kerala, India*
drkpsanthosh@gmail.com

**Abstract.**

The feasibility for the alpha decay and the heavy particle decay from the even-even superheavy (SH) nuclei with Z = 116-124 have been studied within the Coulomb and proximity potential model (CPPM). The Universal formula for cluster decay (UNIV) of Poenaru et al., the Universal Decay Law (UDL) of Qi et al., and the Scaling Law of Horoi et al., has also been used for the evaluation of the decay half lives. A comparison of our predicted half lives with the values evaluated using these empirical formulas are in agreement with each other and hence CPPM could be considered as a unified model for alpha and cluster decay studies. The spontaneous fission half lives of the corresponding parents have also been evaluated using the semi-empirical formula of Santhosh et al. Within our fission model, we have studied cluster formation probability for various clusters and the maximum cluster formation probability for the decay accompanying $^{298}$114 reveals its doubly magic behavior. In the plots for $\log_{10}(T_{1/2})$ against the neutron number of the daughter in the corresponding decay, the half life is found to be the minimum for the decay leading to $^{298}$114 (Z = 114, N = 184) and this also indicate its doubly magic behavior. Most of the predicted half lives are well within the present upper limit for measurements ($T_{1/2} < 10^{30} s$) and the computed alpha half lives for $^{290,292}$116 agrees well with the experimental data. We have thus confidently indicate towards a new island for the cluster radioactivity around the superheavy isotope $^{298}$114 and its neighbors and we hope to receive experimental information about the cluster decay half lives of these considered SHs', hoping to confirm the present calculations.

## 1. Introduction

The radioactive decays, especially the alpha decay and the heavy particle decay, have been a topic of great interest among both the experimentalists and the theoreticians. The exotic decay of particle heavier than the alpha particle, referred to as the cluster decay which comes under the class of cold decays, contradicts the hot fission through its exclusive feature of the formation of the decay products in the ground or the lowest excited states. This well established, rare, cold (neutron-less) mode of decay was first predicted by Sandulescu et al., [1] in 1980 on the basis of quantum mechanical fragmentation theory [2], numerical and analytical superasymmetric fission models, as well as by extending the alpha decay theory to heavier fragments [3]. The first experimental confirmation of such decay was given by Rose and Jones [4] in 1984 through the radioactive decay of $^{223}$Ra by the emission of $^{14}$C. Later, Tretyakova et al., [5] observed the decay of $^{24}$Ne from $^{231}$Pa in Dubna using the solid state nuclear track detectors (SSNTD) and this method is found to be the most effective for cluster decay studies. Extensive experimental search for cluster emission from various parents in the trans-lead region has led to the detection of about 20 cases of spontaneous emission of clusters ranging from $^{14}$C to $^{34}$Si from $^{221}$Fr to $^{242}$Cm [6]. The emission of heavier fragments in such a way that daughter nuclei are always doubly magic or nearly doubly magic (i.e. $^{208}$Pb or closely neighboring nuclei) can be regarded as the farfetched feature of these emissions.

The studies on the competition of α decay and cluster decay in the region of the heaviest superheavy (SH) nuclei have turned out to be a hot topic among both theoreticians and experimentalists, as these studies could provide valuable information regarding the stability, mode of decay and structure of these nuclei. The investigations for the existence of SH nuclei beyond the valley of stability and the urge to reach the island of stability around Z = 120, 124 or 126 and N = 184 [7] have been progressing through a series of cold fusion experiments (performed at GSI, Darmstadt and RIKEN, Japan) [8, 9] and hot fusion reactions (performed at JINR-FLNR (Dubna)) [10]. The successful synthesis of the heavy elements with Z = 107-112 at GSI, Darmstadt [8, 11-13], is to be recalled as the first triumph in the production of SH nuclei. Later on, Oganessian et al., in collaboration with the LLNL researchers [14-18], synthesized the SH nuclei with Z = 113-116 and 118 along with the isotopes of Z = 107-112 at JINR-FLNR, Dubna and very recently they were also successful in the synthesis of two isotopes of Z = 117 [19]. Morita et al., have identified an isotope of Z = 113 at RIKEN, Japan [9, 20] and have also reconfirmed [21] the existence of Z = 110, 111, and 112 reported earlier by GSI group. The majority of proton-rich SH nuclei are identified through the α decay chains and up to now, only α decay, β decay, and spontaneous fission of SH nuclei have been observed. Even though new experiments are presently running at GSI Darmstadt and attempts to produce Z = 120 are reported [22], the heaviest element known so far is Z = 118 [17, 23], and any further progress in the synthesis of new elements with Z > 118 is not quite evident. The low probability of formation, and the separation of the short lived compound nucleus from the very high flux of incident projectile nuclei can be quoted [24] as the main experimental difficulties in identifying the new SH nuclei.

Extensive theoretical studies have been performed on both the cluster decay and alpha decay from heavy and SH nuclei within various theoretical models. A truly universal formula, valid for the radioactivity of all clusters, including α particles was given by Qi et al., [25, 26] on the basis of the microscopic mechanism of the charged particle emission. Since 1984, the Analytical Superasymmetric Fission Model (ASAFM) have been successfully used [27, 28] to compute half life for alpha and cluster radioactivity in heavy and superheavy nuclides. Recently Poenaru et al., [24] have changed the concept of heavy-particle radioactivity (HPR) to allow emitted particles with $Z_e > 28$ from parents with Z > 110 and daughter around $^{208}$Pb and the study revealed the possibility of observing regions in the superheavy nuclei were HPR is stronger than alpha radioactivity. Later, the authors were successful in developing the universal curve [29, 30] for α-decay and cluster radioactivities, based on the fission approach of these decay modes, and the results were compared with the universal decay law for a total of 534 α-emitters in four groups: even-even, even-odd, odd-even and odd-odd. A similar study [31] on the cluster decay of superheavy nuclei performed recently by the authors gave an unexpected result that, for some of the SH nuclei, cluster decay dominates over α decay. A unified formula of half lives for α decay and cluster decay has been given by Ni et al., [32-34] to study the decay of even-even nuclei and their analysis of cluster radioactivity using the new formula have successfully reproduced the cluster decay half lives. The new formula can be considered as a natural extension of Geiger–Nuttall law and Viola–Seaborg formula. The decay properties and the stability of the heavy nuclei with $Z \leq 132$ have been studied by Karpov et al., [35] within the macro-microscopical approach for nuclear ground state masses and phenomenological relations for the half-lives with respect to α-decay, β-decay and spontaneous fission.

Theoretical predictions on the shell closures at Z = 114, 120, 126 and N = 162, 172, 184 in the super-heavy mass region have been done by many authors [36-38]. Within the preformed cluster model, Gupta et al., [38] have calculated the alpha half-life time value of $^{285}$114, indicating that the isotope is stable against α decay and the magicity of protons at Z = 114 or of neutrons at N ≈ 172 was accounted for this stability. Alpha decay studies on Z = 122 [39] and cluster decay studies based on the concept of cold valley in fission and fusion on Z = 116 [40] by Santhosh et al., also indicate neutron shell closure at N = 162, 184 and proton shell closure at Z = 114 and have shown that $^{298}$114 is the spherical doubly magic nuclei. The Coulomb and proximity potential model (CPPM) [41, 42] and its modified version, the Coulomb and proximity potential model for deformed nuclei (CPPMDN) [43] is being used by the Santhosh et al., since the last decade for the cluster decay and alpha decay studies of heavy [44-47] and superheavy nuclei [48-52]. The recent studies on the α transitions from both the ground state and the isomeric states [53], α fine structure studies of even-even [43], even-odd [54], odd-even [55] and odd-odd [56] nuclei, studies on the α decay half lives of the isotopes of Bi nuclei [47] and the theoretical predictions on the alpha decay chains of the isotopes of Z = 115 [49], 117 [50, 51] and 119 [52] have proved the validity and applicability of both CPPM and CPPMDN. The present paper deals with an investigation on the cluster decay of even-even clusters $^{4}$He, $^{8}$Be, $^{10}$Be, $^{14}$C, $^{20}$O and $^{24}$Ne from the various even-even superheavy parent isotopes $^{290-314}$116, $^{294-318}$118, $^{296-320}$118, $^{300-324}$120, $^{306-330}$122 and $^{310-334}$124 leading to the daughter nucleus $^{298}$114 (Z = 114, N = 184) within CPPM. We have considered all the parent-cluster combinations (clusters up to Z = 10), and we would confidently like to mention here that, through our theoretical study we have predicted $^{298}$114 as the next doubly magic nucleus. The values obtained using the Universal Decay Law (UDL) of Qi et al. [30, 31], the Universal (UNIV) curve of Poenaru et al. [34, 35] and the Scaling Law of Horoi et al. [57] have been used for the comparison of our calculations. Since spontaneous fission is another important mode of decay in the superheavy region, we have also computed the spontaneous fission half lives of all the parent nuclei under study

A detail presentation of the Coulomb and Proximity Potential Model (CPPM) is given in Section 2. In Section 3, we have given the results and discussions on the cluster decay of the nuclei under study and the conclusion on the entire work is given in Section 4.

## 2. The coulomb and proximity potential model (CPPM)

For the touching configuration and for the separated fragments, the potential energy barrier in Coulomb and proximity potential model (CPPM) is taken as the sum of Coulomb potential, proximity potential and centrifugal potential. The simple power law interpolation as done by Shi and Swiatecki [58] is used for the pre-scission (overlap) region. Shi and Swiatecki [58] were the first to use the proximity potential in an empirical manner and later on, Gupta et al., [59] have quite extensively used it in the preformed cluster model (PCM), based on pocket formula of Blocki et al., [60] given as:

$$\Phi(\varepsilon) = -\left(\frac{1}{2}\right)(\varepsilon - 2.54)^2 - 0.0852(\varepsilon - 2.54)^3, \text{ for } \varepsilon \leq 1.2511 \quad (1)$$

$$\Phi(\varepsilon) = -3.437 \exp\left(\frac{-\varepsilon}{0.75}\right), \text{ for } \varepsilon \geq 1.2511 \quad (2)$$

where $\Phi$ is the universal proximity potential. The different versions of proximity potentials have been used by R K Puri *et al.*, [61, 62] for studying fusion cross section of different target-projectile combinations. Another formulation of proximity potential [63] is been used in the present model, as given by Eqs. 6 and 7, and the assault frequency $v$ is calculated for each parent-cluster combination which is associated with vibration energy. But, for even A parents and for odd A parents, Shi and Swiatecki [64] get $v$ empirically, unrealistic values as $10^{22}$ and $10^{20}$ respectively.

The interacting potential barrier for a parent nucleus exhibiting cluster decay is given by,

$$V = \frac{Z_1 Z_2 e^2}{r} + V_p(z) + \frac{\hbar^2 \ell(\ell+1)}{2\mu r^2}, \text{ for } z > 0 \quad (3)$$

Here $Z_1$ and $Z_2$ are the atomic numbers of the daughter and emitted cluster, '$z$' is the distance between the near surfaces of the fragments, '$r$' is the distance between fragment centers and is given as $r = z + C_1 + C_2$, where, $C_1$ and $C_2$ are the Süsmann central radii of fragments. The term $\ell$ represents the angular momentum, $\mu$ the reduced mass and $V_P$ is the proximity potential. The proximity potential $V_P$ is given by Blocki et al. [60] as,

$$V_p(z) = 4\pi\gamma b \left[\frac{C_1 C_2}{(C_1 + C_2)}\right] \Phi\left(\frac{z}{b}\right), \quad (4)$$

with the nuclear surface tension coefficient,

$$\gamma = 0.9517[1 - 1.7826(N-Z)^2 / A^2] \text{ MeV/fm}^2 \quad (5)$$

where $N$, $Z$ and $A$ represent neutron, proton and mass number of parent respectively, $\Phi$ represents the universal the proximity potential [63] given as

$$\Phi(\varepsilon) = -4.41 e^{-\varepsilon/0.7176}, \text{ for } \varepsilon > 1.9475 \quad (6)$$

$$\Phi(\varepsilon) = -1.7817 + 0.9270\varepsilon + 0.0169\varepsilon^2 - 0.05148\varepsilon^3, \text{ for } 0 \leq \varepsilon \leq 1.9475 \quad (7)$$

With $\varepsilon = z/b$, where the width (diffuseness) of the nuclear surface $b \approx 1$ and Süsmann central radii $C_i$ of fragments related to sharp radii $R_i$ as,

$$C_i = R_i - \left(\frac{b^2}{R_i}\right) \quad (8)$$

For $R_i$ we use semi empirical formula in terms of mass number $A_i$ as [60],

$$R_i = 1.28 A_i^{1/3} - 0.76 + 0.8 A_i^{-1/3} \tag{9}$$

The potential for the internal part (overlap region) of the barrier is given as,

$$V = a_0 (L - L_0)^n, \text{ for } z < 0 \tag{10}$$

Here $L = z + 2C_1 + 2C_2$ and $L_0 = 2C$, the diameter of the parent nuclei. The constants $a_0$ and $n$ are determined by the smooth matching of the two potentials at the touching point.

Using one dimensional WKB approximation, the barrier penetrability P is given as,

$$P = \exp\left\{-\frac{2}{\hbar}\int_a^b \sqrt{2\mu(V-Q)}\,dz\right\} \tag{11}$$

Here the mass parameter is replaced by $\mu = mA_1 A_2 / A$, where 'm' is the nucleon mass and $A_1$, $A_2$ are the mass numbers of daughter and emitted cluster respectively. The turning points "a" and "b" are determined from the equation $V(a) = V(b) = Q$. The above integral can be evaluated numerically or analytically, and the half life time is given by

$$T_{1/2} = \left(\frac{\ln 2}{\lambda}\right) = \left(\frac{\ln 2}{\nu P}\right) \tag{12}$$

where, $\nu = \left(\frac{\omega}{2\pi}\right) = \left(\frac{2E_v}{h}\right)$ represent the number of assaults on the barrier per second and $\lambda$ the decay constant. $E_v$, the empirical vibration energy is given as [65],

$$E_v = Q\left\{0.056 + 0.039 \exp\left[\frac{(4-A_2)}{2.5}\right]\right\}, \quad \text{for } A_2 \geq 4 \tag{13}$$

Within our fission model (CPPM) the cluster formation probability $S$ can be calculated as the penetration of the internal part (overlap region) of the barrier given as

$$S = \exp(-K) \tag{14}$$

where

$$K = \frac{2}{\hbar}\int_a^0 \sqrt{2\mu(V-Q)}\,dz \tag{15}$$

here, $a$ is the inner turning point and is defined as $V(a) = Q$ and $z = 0$ represent the touching configuration.

In the present model, we have included the probability of formation of the cluster before its emission. The decay constant $\lambda$ and the penetrability through the total potential barrier $P$ is related as, $\lambda = \nu P$, where $\nu$ is the assault frequency and $P = SP^{ext.}$. Within a fission model, the cluster formation probability $S$ can be calculated as the penetrability through the internal part (overlap region) of the barrier and is given in equation (14) and (15). $P^{ext.}$ is the penetrability through the external potential barrier and is given as

$$P = \exp\left\{-\frac{2}{\hbar}\int_0^b \sqrt{2\mu(V-Q)}\,dz\right\} \tag{16}$$

The first turning point z = 0, represents the touching configuration and z = b represents the outer turning point which can be determined using the equation V(b) = Q.

## 3. Results and discussions

The decay half lives in the emission of even-even clusters $^4$He, $^8$Be, $^{10}$Be, $^{14}$C, $^{20}$O and $^{24}$Ne from the various even-even superheavy parent isotopes $^{290-314}$116, $^{294-318}$118, $^{296-320}$118, $^{300-324}$120, $^{306-330}$122 and $^{310-334}$124 leading to the predicted [36-40] doubly magic $^{298}$114 (Z = 114, N = 184) and the neighboring nuclei have been calculated by using the Coulomb and proximity potential model (CPPM). The possibility to have a cluster decay process is related to its exotermicity, Q > 0. The energy released in decay transitions between the ground state energy levels of the parent nuclei and the ground state energy levels of the daughter nuclei is given as

$$Q_{gs \to gs} = \Delta M_p - (\Delta M_c + \Delta M_d) + k(Z_p^\varepsilon - Z_d^\varepsilon) \tag{17}$$

where $\Delta M_p$, $\Delta M_d$, $\Delta M_c$ are the mass excess of the parent, daughter and cluster respectively. The Q values for cluster decays are calculated using the experimental mass excess values of Wang *et al.*, [66] and some of the mass excess were taken from Koura-Tachibana-Uno-Yamada (KTUY) [67], as those experimental mass excess were unavailable in Ref [66]. As the effect of atomic electrons on the energy of the cluster has not been included in Ref. [66, 67], for a more accurate calculation of Q value, we have included the electron screening effect [68] in equation (17). The term $k(Z_p^\varepsilon - Z_d^\varepsilon)$ represents this correction, where the quantity $kZ^\varepsilon$ represents the total binding energy of the Z electrons in the atom. Here the values of $k$ = 8.7eV and $\varepsilon$ = 2.517 for nuclei with Z ≥ 60; and $k$ = 13.6eV and $\varepsilon$ = 2.408 for nuclei with Z < 60, have been derived from data reported by Huang et al., [69].

### 3.1 Alpha decay half lives

The alpha decay half lives for the isotopes under study have also been evaluated within the Universal Decay Law (UDL) of Qi et al., [30, 31], the Universal (UNIV) curve of Poenaru et al., [34, 35] and the Scaling Law of Horoi et al., [57]. The formalisms are discussed below.

### 3.1. 1 The Universal Curve (UNIV)

The decay half lives have been explained using several simple and effective relationships, which are obtained by fitting the experimental data. The universal (UNIV) curves [70-73], derived by extending a fission theory to larger mass asymmetry should be mentioned, among them, with great importance. Based on the quantum mechanical tunnelling process [74, 75], in UNIV, the disintegration constant λ, valid in both fission-like and α-like theories and the partial decay half life T of the parent nucleus is related as,

$$\lambda = \ln 2/T = \nu S P_s \tag{18}$$

Here *v, S* and $P_s$ are three model-dependent quantities: *v* is the frequency of assaults on the barrier per second, *S* is the pre-formation probability of the cluster at the nuclear surface (equal to the penetrability of the internal part of the barrier in a fission theory [70, 71]), and $P_s$ is the quantum penetrability of the external potential barrier.

By using the decimal logarithm,

$$\log_{10} T(s) = -\log_{10} P - \log_{10} S + [\log_{10}(\ln 2) - \log_{10} \nu] \tag{19}$$

To derive the universal formula, it was assumed that $\nu$ = constant and that $S$ depends only on the mass number of the emitted particle $A_e$ [71, 74] as the microscopic calculation of the pre-formation probability [76] of many clusters from $^8$Be to $^{46}$Ar had shown that it is dependent only upon the size of the cluster. Using a fit with experimental data for $\alpha$ decay, the corresponding numerical values [71] obtained were, $S_\alpha$ = 0.0143153, $\nu$ = $10^{22.01} s^{-1}$. The decimal logarithm of the pre-formation factor is given as,

$$\log_{10} S = -0.598(A_e - 1) \tag{20}$$

and the additive constant for an even-even nucleus is,

$$c_{ee} = [-\log_{10} \nu + \log_{10}(\ln 2)] = -22.16917 \tag{21}$$

The penetrability of an external Coulomb barrier, having separation distance at the touching configuration $R_a = R_t = R_d + R_e$ as the first turning point and the second turning point defined by $e^2 Z_d Z_e / R_b = Q$, may be found analytically as

$$-\log_{10} P_S = 0.22873(\mu_A Z_d Z_e R_b)^{1/2} \times [\arccos\sqrt{r} - \sqrt{r(1-r)}] \tag{22}$$

where $r = R_t / R_b$, $R_t = 1.2249(A_d^{1/3} + A_e^{1/3})$ and $R_b = 1.43998 Z_d Z_e / Q$

The released energy $Q$ is evaluated using the mass tables [66] and the liquid-drop-model radius constant $r_0$ = 1.2249fm.

### 3.1.2 The Universal Decay Law (UDL)

Starting from the $\alpha$-like (extension to the heavier cluster of $\alpha$-decay theory) $R$-matrix theory and the microscopic mechanism of the charged-particle emission, a new universal decay law (UDL) for $\alpha$-decay and cluster decay modes was introduced [25, 26] by Qi et al.,. The model was presented in an interesting way, which made it possible to represent, on the same plot with a single straight line, the logarithm of the half lives minus some quantity versus one of the two parameters ($\chi'$ and $\rho'$) that depend on the atomic and mass numbers of the daughter and emitted particles as well as the $Q$ value. UDL relates the half-life of monopole radioactive decay with the $Q$ values of the outgoing particles as well as the masses and charges of the nuclei involved in the decay and can be written in the logarithmic form as,

$$\log_{10}(T_{1/2}) = a Z_c Z_d \sqrt{\frac{A}{Q_c}} + b\sqrt{A Z_c Z_d (A_d^{1/3} + A_c^{1/3})} + c \tag{23}$$

$$= a\chi' + b\rho' + c \tag{24}$$

where the quantity $A = \dfrac{A_d A_c}{A_d + A_c}$ and the constants $a$ = 0.4314, $b$ = -0.4087 and $c$ = -25.7725 are the coefficient sets of eq. (23), determined by fitting to experiments of both $\alpha$ and cluster decays [25].

The effects that induce the clusterization in the parent nucleus are included in the term $b\rho'+c$. As this relation holds for the monopole radioactive decays of all clusters, it is called the Universal Decay Law (UDL) [25].

### 3.1.3 Scaling law of Horoi et al.,

In order to determine the half lives of both the alpha and cluster decays, a new empirical formula for cluster decay was introduced by Horoi et al., [57] and is given by the equation,

$$\log_{10} T_{1/2} = (a_1\mu^x + b_1)[(Z_1 Z_2)^y / \sqrt{Q} - 7] + (a_2\mu^x + b_2) \quad (25)$$

where $\mu$ is the reduced mass. The six parameters are $a_1 = 9.1$, $b_1 = -10.2$, $a_2 = 7.39$, $b_2 = -23.2$, $x = 0.416$ and $y = 0.613$.

## 3.2. Spontaneous fission half lives

Spontaneous fission, the limiting factor that determines the stability of newly synthesized super heavy nuclei, may be considered as one of the most prominent decay modes, energetically feasible for both heavy and superheavy nuclei with proton number $Z \geq 90$. The spontaneous fission half lives of the parent isotopes under study have been evaluated using the semi-empirical relation of Santhosh et al., [77] discussed below.

### 3.2.1 Semi-empirical relation of Santhosh et al.,

A new semi empirical formula for explaining spontaneous fission was developed by Santhosh et al., [77] by making least squares fit to the available experimental data. The formula obtained for logarithmic half-life time for spontaneous fission is given by

$$\log_{10}(T_{1/2}/yr) = a\frac{Z^2}{A} + b\left(\frac{Z^2}{A}\right)^2 + c\left(\frac{N-Z}{N+Z}\right) + d\left(\frac{N-Z}{N+Z}\right)^2 + e, \quad (26)$$

where, the constants are $a = -43.25203$, $b = 0.49192$, $c = 3674.3927$, $d = -9360.6$ and $e = 580.75058$. Here the quantities $\frac{Z^2}{A}$ and $I = \frac{N-Z}{N+Z}$ are the fissionability parameter and the neutron excess of the decaying parent nuclei respectively. It is to be noted that the semi-empirical formula works well for the nuclei in the mass regions $^{232}$Th to $^{286}$114 [77].

Figures 1-3 represent the plot for $\log_{10}(S)$ vs. neutron number of the parent nuclei, for the cluster emission of $^4$He, $^8$Be, $^{10}$Be, $^{14}$C, $^{20}$O and $^{24}$Ne respectively from $^{290-314}$116, $^{294-318}$118, $^{296-320}$118, $^{300-324}$120, $^{306-330}$122, $^{310-334}$124. The behavior of the cluster formation probability with the neutron number of the parent nuclei can be clearly seen from these figures. In figure 1 (a) and 1 (b), the plot for the cluster formation probability of $^4$He from $^{290-314}$116 and $^8$Be from $^{294-318}$118 isotopes have been given and it is to be noticed that the cluster formation probability is the maximum for the emission of $^4$He and $^8$Be accompanied by $^{298}$114 (Z = 114, N = 184) daughter nuclei. The plots for the cluster formation probability of $^{10}$Be from $^{296-320}$118 and $^{14}$C from $^{300-324}$120 isotopes have been given respectively in figure 2 (a) and 2 (b). It should be noticed that the cluster formation probability

is the maximum for the emission of $^{10}$Be and $^{14}$C accompanied by $^{298}$114 (N = 84, Z = 114) daughter nuclei. In figure 3 (a) and 3 (b), the plot for the cluster formation probability of $^{20}$O from $^{306-330}$122 and $^{24}$Ne from $^{310-334}$124 isotopes have been given and it can be clearly seen that the cluster formation probability is the maximum for the emission of $^{20}$O and $^{24}$Ne accompanied by $^{298}$114 (Z = 114, N = 184) daughter nuclei. Thus it is clearly evidenced from the figures 1-3 that, the cluster formation probability is maximum for the decay accompanying $^{298}$114 and this reveal the doubly magic behavior of $^{298}$114.

The cluster decay half lives have been evaluated using CPPM, UNIV, UDL and the scaling law of Horoi and their comparisons are shown in figures 4-6. The plots for $\log_{10}(T_{1/2})$ against the neutron number of the daughter in the corresponding decay are given in these figures. Fig. 4 gives the plot for the cluster emission of $^{4}$He and $^{8}$Be from $^{290-314}$116 and $^{294-318}$118 isotopes respectively. In fig. 5 and fig. 6, the plots for the cluster emission of $^{10}$Be from $^{296-320}$118, $^{14}$C from $^{300-324}$120 and $^{20}$O from $^{306-330}$122, $^{24}$Ne from $^{310-334}$124 isotopes have been given respectively. The minima of the logarithmic half-lives for all these cluster emission are found for the decay leading to $^{298}$114 (Z = 114, N = 184). A minimum in the decay half lives corresponds to the greater barrier penetrability, which in turn indicates the doubly magic behavior of the daughter nuclei. In the cluster decay studies on heavy nuclei, it has been shown that the half life is minimum for the decays leading to the doubly magic daughter $^{208}$Pb (Z = 82, N = 126) or its neighboring nuclei. The present study on the cluster decay half lives of the superheavy nuclei gives a pronounced minima for the daughter $^{298}$114 (Z = 114, N = 184). This may be interpreted as a result of the strong shell effect of the assumed magic number of the neutrons and protons and this reveal the role the doubly magic $^{298}$114 in cluster decays of superheavy nuclei.

It can be also be seen from the plots connecting $\log_{10}(T_{1/2})$ versus neutron number of daughter nuclei that the four calculations, CPPM, UNIV, UDL and Scaling law, show the same trend. It should be taken into consideration that the CPPM values matches well with the UDL values than that of the UNIV or the values obtained using the Scaling Law of Horoi. Thus, similar to UNIV, UDL and Scaling law, CPPM could be considered as a unified model for α-decay and cluster decay studies.

In Tables 1-3, the computed Q values, barrier penetrability, decay constant and half-lives for the emission of various cluster from the superheavy nuclei $^{290-314}$116, $^{294-318}$118, $^{296-320}$118, $^{300-324}$120, $^{306-330}$122, $^{310-334}$124 are given. The parent nuclei, the emitted clusters and the corresponding daughter nuclei are given in columns 1, 2 and 3 respectively of the tables mentioned above. Column 4 gives the respective Q values of these decays which are evaluated using Eq. (17). The penetrability and decay constants for the respective decays are calculated using CPPM and are included in columns 5 and 6 respectively. The cluster decay half-lives predicted within the CPPM for all the parent-cluster combinations are arranged in column 7. Most of the predicted half lives are well within the present upper limit for measurements $(T_{1/2} < 10^{30} s)$. Moreover, the alpha half lives calculated using our model give closer values with the experimental alpha half lives [78]. For example, in the case of $^{290}$116, the $T_{\alpha}^{\exp} = 1.500 \times 10^{-2} s$ and $T_{\alpha}^{calc.}(CPPM) = 5.259 \times 10^{-2} s$ and in the case of $^{292}$116, the $T_{\alpha}^{\exp} = 1.800 \times 10^{-2} s$ and $T_{\alpha}^{calc.}(CPPM) = 1.951 \times 10^{-1} s$. Spontaneous fission, being an important mode of decay in the superheavy region, we have computed the spontaneous fission half lives of all the parent nuclei under study, using the semi-empirical formula of Santhosh et al., [77] and the corresponding values have been given in the columns 8.

Thus the present study on the cluster decay half lives for the emission of various clusters from the superheavy nuclei $^{290-314}$116, $^{294-318}$118, $^{296-320}$118, $^{300-324}$120, $^{306-330}$122, $^{310-334}$124 reveals that the cluster decay half lives is the minimum for those decays leading to the daughter nuclei $^{298}$114 with Z = 114 and N = 184, the next predicted proton and neutron shell closures. So through our study, we could confidently predict the new island for the cluster radioactivity leading to the residual superheavy isotope $^{298}$114 and its neighbors. We would like to mention that, the results obtained through our study closely agree with that of the early predictions [36-40]. Thus we have established the fact that, the isotope $^{298}$114 should be considered as the next predicted spherical doubly magic nucleus after the experimentally observed doubly magic nuclei $^{208}$Pb and $^{100}$Sn.

## 4. Conclusion

Calculations on the cluster decay half lives for the emission of $^4$He, $^8$Be, $^{10}$Be, $^{14}$C, $^{20}$O and $^{24}$Ne from the various superheavy parents $^{290-314}$116, $^{294-318}$118, $^{296-320}$118, $^{300-324}$120, $^{306-330}$122 and $^{310-334}$124 leading to the predicted doubly magic $^{298}$114 (Z = 114, N = 184) and the neighboring nuclei have been by taking the barrier potential as the sum of Coulomb and proximity potential (within CPPM). A comparison of our calculated alpha and cluster half lives with that of the values evaluated within the Universal formula for cluster decay (UNIV) of Poenaru et al., the Universal Decay Law (UDL) and the Scaling Law of Horoi et al. show a similar trend. The spontaneous fission half lives of the corresponding parents have also been evaluated using the semi-empirical formula of Santhosh et al.,. The behavior of the cluster formation probability with the neutron number of the parent nuclei can be clearly seen from the plots for $\log_{10}(S)$ vs. neutron number of the parent nuclei. The role of neutron magicity in cluster decays is clearly revealed from the low values of the cluster decay half-lives at N = 184, as seen in the plots for $\log_{10}(T_{1/2})$ versus neutron number of daughter nuclei. We have thus established the fact that, the isotope $^{298}$114 should be considered as the next predicted spherical doubly magic nucleus and thus our study indicate towards a new island for the cluster radioactivity leading to the residual superheavy isotope $^{298}$114 and its neighbors.

**Acknowledgments**

The author KPS would like to thank the University Grants Commission, Govt. of India for the financial support under Major Research Project. No.42-760/2013 (SR) dated 22-03-2013.

**Table 1.** The Q value, penetrability, decay constant and the predicted half lives for the emission of the cluster $^4$He from $^{290-314}$116 isotopes and the cluster $^8$Be from $^{294-318}$118 isotopes. The half lives are calculated for zero angular momentum transfers.

| Parent nuclei | Emitted cluster | Daughter nuclei | Q value (MeV) | Penetrability P | Decay constant $\lambda$ (s$^{-1}$) | $T^{\alpha}_{1/2}$ (s) CPPM | $T^{sf}_{1/2}$ (s) KPS [77] |
|---|---|---|---|---|---|---|---|
| $^{290}$116 | $^4$He | $^{286}$114 | 11.054 | 2.595x10$^{-20}$ | 1.318x10$^1$ | 5.259x10$^{-2}$ | 7.831x10$^0$ |
| $^{292}$116 | $^4$He | $^{288}$114 | 10.834 | 7.136x10$^{-21}$ | 3.552x10$^0$ | 1.951x10$^{-1}$ | 3.293x10$^{-1}$ |
| $^{294}$116 | $^4$He | $^{290}$114 | 10.224 | 1.395x10$^{-22}$ | 6.553x10$^{-2}$ | 1.058x10$^1$ | 5.090x10$^{-3}$ |
| $^{296}$116 | $^4$He | $^{292}$114 | 10.564 | 1.457x10$^{-21}$ | 7.073x10$^{-1}$ | 9.798x10$^{-1}$ | 3.015x10$^{-5}$ |
| $^{298}$116 | $^4$He | $^{294}$114 | 10.324 | 3.171x10$^{-22}$ | 1.504x10$^{-1}$ | 4.608x10$^0$ | 7.117x10$^{-8}$ |
| $^{300}$116 | $^4$He | $^{296}$114 | 10.194 | 1.398x10$^{-22}$ | 6.549x10$^{-2}$ | 1.058x10$^1$ | 6.957x10$^{-11}$ |
| $^{302}$116 | $^4$He | $^{298}$114 | 11.784 | 2.772x10$^{-18}$ | 1.501x10$^3$ | 4.617x10$^{-4}$ | 2.921x10$^{-14}$ |
| $^{304}$116 | $^4$He | $^{300}$114 | 10.944 | 2.135x10$^{-20}$ | 1.073x10$^1$ | 6.457x10$^{-2}$ | 5.453x10$^{-18}$ |
| $^{306}$116 | $^4$He | $^{302}$114 | 10.184 | 1.586x10$^{-22}$ | 7.420x10$^{-2}$ | 9.339x10$^0$ | 4.684x10$^{-22}$ |
| $^{308}$116 | $^4$He | $^{304}$114 | 9.424 | 6.597x10$^{-25}$ | 2.856x10$^{-4}$ | 2.426x10$^3$ | 1.912x10$^{-26}$ |
| $^{310}$116 | $^4$He | $^{306}$114 | 8.474 | 2.494x10$^{-28}$ | 9.708x10$^{-8}$ | 7.139x10$^6$ | 3.824x10$^{-31}$ |
| $^{312}$116 | $^4$He | $^{308}$114 | 8.214 | 2.384x10$^{-29}$ | 8.998x10$^{-9}$ | 7.702x10$^7$ | 3.862x10$^{-36}$ |
| $^{314}$116 | $^4$He | $^{310}$114 | 7.984 | 2.730x10$^{-30}$ | 1.001x10$^{-9}$ | 6.920x10$^8$ | 2.027x10$^{-41}$ |
| $^{294}$118 | $^8$Be | $^{286}$114 | 22.837 | 1.022x10$^{-39}$ | 7.209x10$^{-19}$ | 9.613x10$^{17}$ | 5.915x10$^3$ |
| $^{296}$118 | $^8$Be | $^{288}$114 | 20.927 | 8.592x10$^{-45}$ | 5.554x10$^{-24}$ | 1.248x10$^{23}$ | 2.369x10$^2$ |
| $^{298}$118 | $^8$Be | $^{290}$114 | 21.307 | 1.192x10$^{-43}$ | 7.843x10$^{-23}$ | 8.836x10$^{21}$ | 3.590x10$^0$ |
| $^{300}$118 | $^8$Be | $^{292}$114 | 21.567 | 7.256x10$^{-43}$ | 4.834x10$^{-22}$ | 1.434x10$^{21}$ | 2.140x10$^{-2}$ |
| $^{302}$118 | $^8$Be | $^{294}$114 | 21.207 | 8.370x10$^{-44}$ | 5.483x10$^{-23}$ | 1.264x10$^{22}$ | 5.217x10$^{-5}$ |
| $^{304}$118 | $^8$Be | $^{296}$114 | 22.597 | 5.225x10$^{-40}$ | 3.647x10$^{-19}$ | 1.900x10$^{18}$ | 5.392x10$^{-8}$ |
| $^{306}$118 | $^8$Be | $^{298}$114 | 23.647 | 2.421x10$^{-37}$ | 1.769x10$^{-16}$ | 3.918x10$^{15}$ | 2.447x10$^{-11}$ |
| $^{308}$118 | $^8$Be | $^{300}$114 | 22.207 | 6.660x10$^{-41}$ | 4.568x10$^{-20}$ | 1.517x10$^{19}$ | 5.045x10$^{-15}$ |
| $^{310}$118 | $^8$Be | $^{302}$114 | 20.427 | 8.049x10$^{-46}$ | 5.079x10$^{-25}$ | 1.365x10$^{24}$ | 4.879x10$^{-19}$ |
| $^{312}$118 | $^8$Be | $^{304}$114 | 18.667 | 2.370x10$^{-51}$ | 1.367x10$^{-30}$ | 5.070x10$^{29}$ | 2.284x10$^{-23}$ |
| $^{314}$118 | $^8$Be | $^{306}$114 | 17.477 | 1.552x10$^{-55}$ | 8.377x10$^{-35}$ | 8.272x10$^{33}$ | 5.331x10$^{-28}$ |
| $^{316}$118 | $^8$Be | $^{308}$114 | 16.997 | 2.585x10$^{-57}$ | 1.357x10$^{-36}$ | 5.106x10$^{35}$ | 6.385x10$^{-33}$ |
| $^{318}$118 | $^8$Be | $^{310}$114 | 16.527 | 3.936x10$^{-59}$ | 2.010x10$^{-38}$ | 3.447x10$^{37}$ | 4.035x10$^{-38}$ |

**Table 2.** The Q value, penetrability, decay constant and the predicted half lives for the emission of the cluster $^{10}$Be from $^{296-320}$118 isotopes and the cluster $^{14}$C from $^{300-324}$120 isotopes. The half lives are calculated for zero angular momentum transfers.

| Parent nuclei | Emitted cluster | Daughter nuclei | Q value (MeV) | Penetrability P | Decay constant $\lambda$ (s$^{-1}$) | $T_{1/2}^{\alpha}$ (s) CPPM | $T_{1/2}^{sf}$ (s) KPS [77] |
|---|---|---|---|---|---|---|---|
| $^{296}$118 | $^{10}$Be | $^{286}$114 | 16.371 | 2.682x10$^{-65}$ | 1.264x10$^{-44}$ | 5.481x10$^{43}$ | 2.369x10$^{2}$ |
| $^{298}$118 | $^{10}$Be | $^{288}$114 | 15.931 | 2.635x10$^{-67}$ | 1.209x10$^{-46}$ | 5.733x10$^{45}$ | 3.590x10$^{0}$ |
| $^{300}$118 | $^{10}$Be | $^{290}$114 | 16.831 | 4.450x10$^{-63}$ | 2.156x10$^{-42}$ | 3.214x10$^{41}$ | 2.140x10$^{-2}$ |
| $^{302}$118 | $^{10}$Be | $^{292}$114 | 17.651 | 1.690x10$^{-59}$ | 8.591x10$^{-39}$ | 8.066x10$^{37}$ | 5.217x10$^{-5}$ |
| $^{304}$118 | $^{10}$Be | $^{294}$114 | 19.421 | 1.311x10$^{-52}$ | 7.332x10$^{-32}$ | 9.451x10$^{30}$ | 5.392x10$^{-8}$ |
| $^{306}$118 | $^{10}$Be | $^{296}$114 | 20.791 | 7.359x10$^{-48}$ | 4.406x10$^{-27}$ | 1.573x10$^{26}$ | 2.447x10$^{-11}$ |
| $^{308}$118 | $^{10}$Be | $^{298}$114 | 21.911 | 2.737x10$^{-44}$ | 1.726x10$^{-23}$ | 4.014x10$^{22}$ | 5.045x10$^{-15}$ |
| $^{310}$118 | $^{10}$Be | $^{300}$114 | 20.321 | 2.692x10$^{-49}$ | 1.575x10$^{-28}$ | 4.399x10$^{27}$ | 4.879x10$^{-19}$ |
| $^{312}$118 | $^{10}$Be | $^{302}$114 | 18.341 | 2.028x10$^{-56}$ | 1.071x10$^{-35}$ | 6.470x10$^{34}$ | 2.284x10$^{-23}$ |
| $^{314}$118 | $^{10}$Be | $^{304}$114 | 16.961 | 4.444x10$^{-62}$ | 2.170x10$^{-41}$ | 3.193x10$^{40}$ | 5.331x10$^{-28}$ |
| $^{316}$118 | $^{10}$Be | $^{306}$114 | 16.141 | 9.339x10$^{-66}$ | 4.340x10$^{-45}$ | 1.597x10$^{44}$ | 6.385x10$^{-33}$ |
| $^{318}$118 | $^{10}$Be | $^{308}$114 | 16.011 | 2.551x10$^{-66}$ | 1.176x10$^{-45}$ | 5.893x10$^{44}$ | 4.035x10$^{-38}$ |
| $^{320}$118 | $^{10}$Be | $^{310}$114 | 18.651 | 5.270x10$^{-55}$ | 2.830x10$^{-34}$ | 2.449x10$^{33}$ | 1.381x10$^{-43}$ |
| $^{300}$120 | $^{14}$C | $^{286}$114 | 40.330 | 1.261x10$^{-48}$ | 1.395x10$^{-27}$ | 4.969x10$^{26}$ | 9.776x10$^{5}$ |
| $^{302}$120 | $^{14}$C | $^{288}$114 | 39.800 | 1.090x10$^{-49}$ | 1.190x10$^{-28}$ | 5.824x10$^{27}$ | 1.394x10$^{4}$ |
| $^{304}$120 | $^{14}$C | $^{290}$114 | 40.420 | 3.171x10$^{-48}$ | 3.516x10$^{-27}$ | 1.971x10$^{26}$ | 8.038x10$^{1}$ |
| $^{306}$120 | $^{14}$C | $^{292}$114 | 42.950 | 7.627x10$^{-43}$ | 8.985x10$^{-22}$ | 7.713x10$^{20}$ | 1.943x10$^{-1}$ |
| $^{308}$120 | $^{14}$C | $^{294}$114 | 44.270 | 3.749x10$^{-40}$ | 4.552x10$^{-19}$ | 1.523x10$^{18}$ | 2.038x10$^{-4}$ |
| $^{310}$120 | $^{14}$C | $^{296}$114 | 45.320 | 4.587x10$^{-38}$ | 5.702x10$^{-17}$ | 1.215x10$^{16}$ | 9.605x10$^{-8}$ |
| $^{312}$120 | $^{14}$C | $^{298}$114 | 45.790 | 4.232x10$^{-37}$ | 5.315x10$^{-16}$ | 1.304x10$^{15}$ | 2.099x10$^{-11}$ |
| $^{314}$120 | $^{14}$C | $^{300}$114 | 42.520 | 2.424x10$^{-43}$ | 2.827x10$^{-22}$ | 2.452x10$^{21}$ | 2.195x10$^{-15}$ |
| $^{316}$120 | $^{14}$C | $^{302}$114 | 40.330 | 7.254x10$^{-48}$ | 8.024x10$^{-27}$ | 8.637x10$^{25}$ | 1.132x10$^{-19}$ |
| $^{318}$120 | $^{14}$C | $^{304}$114 | 38.720 | 2.158x10$^{-51}$ | 2.291x10$^{-30}$ | 3.024x10$^{29}$ | 2.962x10$^{-24}$ |
| $^{320}$120 | $^{14}$C | $^{306}$114 | 37.670 | 8.882x10$^{-54}$ | 9.177x10$^{-33}$ | 7.552x10$^{31}$ | 4.043x10$^{-29}$ |
| $^{322}$120 | $^{14}$C | $^{308}$114 | 39.860 | 1.189x10$^{-48}$ | 1.300x10$^{-27}$ | 5.333x10$^{26}$ | 2.957x10$^{-34}$ |
| $^{324}$120 | $^{14}$C | $^{310}$114 | 40.040 | 3.561x10$^{-48}$ | 3.911x10$^{-27}$ | 1.772x10$^{26}$ | 1.188x10$^{-39}$ |

**Table 3.** The Q value, penetrability, decay constant and the predicted half lives for the emission of the cluster $^{20}$O from $^{306-330}$122 isotopes and the cluster $^{24}$Ne from $^{310-334}$124 isotopes. The half lives are calculated for zero angular momentum transfers.

| Parent nuclei | Emitted cluster | Daughter nuclei | Q value (MeV) | Penetrability P | Decay constant $\lambda$ (s$^{-1}$) | $T^{\alpha}_{1/2}$(s) CPPM | $T^{sf}_{1/2}$(s) KPS [77] |
|---|---|---|---|---|---|---|---|
| $^{306}$122 | $^{20}$O | $^{286}$114 | 57.317 | 1.575x10$^{-59}$ | 2.448x10$^{-38}$ | 2.831x10$^{37}$ | 3.088x10$^{8}$ |
| $^{308}$122 | $^{20}$O | $^{288}$114 | 58.847 | 2.983x10$^{-56}$ | 4.759x10$^{-35}$ | 1.456x10$^{34}$ | 1.656x10$^{6}$ |
| $^{310}$122 | $^{20}$O | $^{290}$114 | 61.417 | 4.567x10$^{-51}$ | 7.605x10$^{-30}$ | 9.112x10$^{28}$ | 3.817x10$^{3}$ |
| $^{312}$122 | $^{20}$O | $^{292}$114 | 63.587 | 7.282x10$^{-47}$ | 1.256x10$^{-25}$ | 5.520x10$^{24}$ | 3.912x10$^{0}$ |
| $^{314}$122 | $^{20}$O | $^{294}$114 | 64.137 | 1.000x10$^{-45}$ | 1.739x10$^{-24}$ | 3.985x10$^{23}$ | 1.841x10$^{-3}$ |
| $^{316}$122 | $^{20}$O | $^{296}$114 | 64.347 | 3.240x10$^{-45}$ | 5.653x10$^{-24}$ | 1.226x10$^{23}$ | 4.103x10$^{-7}$ |
| $^{318}$122 | $^{20}$O | $^{298}$114 | 64.477 | 7.446x10$^{-45}$ | 1.302x10$^{-23}$ | 5.324x10$^{22}$ | 4.463x10$^{-11}$ |
| $^{320}$122 | $^{20}$O | $^{300}$114 | 60.087 | 5.248x10$^{-53}$ | 8.550x10$^{-32}$ | 8.105x10$^{30}$ | 2.439x10$^{-15}$ |
| $^{322}$122 | $^{20}$O | $^{302}$114 | 58.037 | 5.226x10$^{-57}$ | 8.224x10$^{-36}$ | 8.427x10$^{34}$ | 6.884x10$^{-20}$ |
| $^{324}$122 | $^{20}$O | $^{304}$114 | 58.807 | 2.520x10$^{-55}$ | 4.018x10$^{-34}$ | 1.725x10$^{33}$ | 1.030x10$^{-24}$ |
| $^{326}$122 | $^{20}$O | $^{306}$114 | 58.457 | 6.357x10$^{-56}$ | 1.008x10$^{-34}$ | 6.878x10$^{33}$ | 8.393x10$^{-30}$ |
| $^{328}$122 | $^{20}$O | $^{308}$114 | 58.977 | 9.225x10$^{-55}$ | 1.475x10$^{-33}$ | 4.698x10$^{32}$ | 3.813x10$^{-35}$ |
| $^{330}$122 | $^{20}$O | $^{310}$114 | 59.307 | 5.413x10$^{-54}$ | 8.704x10$^{-33}$ | 7.962x10$^{31}$ | 9.895x10$^{-41}$ |
| $^{310}$124 | $^{24}$Ne | $^{286}$114 | 84.800 | 9.910x10$^{-52}$ | 2.276x10$^{-30}$ | 3.044x10$^{29}$ | 4.029x10$^{13}$ |
| $^{312}$124 | $^{24}$Ne | $^{288}$114 | 86.140 | 2.611x10$^{-49}$ | 6.093x10$^{-28}$ | 1.137x10$^{27}$ | 1.937x10$^{11}$ |
| $^{314}$124 | $^{24}$Ne | $^{290}$114 | 87.980 | 4.136x10$^{-46}$ | 9.858x10$^{-25}$ | 7.030x10$^{23}$ | 4.102x10$^{8}$ |
| $^{316}$124 | $^{24}$Ne | $^{292}$114 | 89.100 | 4.032x10$^{-44}$ | 9.733x10$^{-23}$ | 7.120x10$^{21}$ | 3.956x10$^{5}$ |
| $^{318}$124 | $^{24}$Ne | $^{294}$114 | 89.200 | 8.736x10$^{-44}$ | 2.111x10$^{-22}$ | 3.283x10$^{21}$ | 1.791x10$^{2}$ |
| $^{320}$124 | $^{24}$Ne | $^{296}$114 | 89.020 | 6.640x10$^{-44}$ | 1.601x10$^{-22}$ | 4.328x10$^{21}$ | 3.922x10$^{-2}$ |
| $^{322}$124 | $^{24}$Ne | $^{298}$114 | 88.880 | 5.788x10$^{-44}$ | 1.394x10$^{-22}$ | 4.973x10$^{21}$ | 4.276x10$^{-6}$ |
| $^{324}$124 | $^{24}$Ne | $^{300}$114 | 86.560 | 1.364x10$^{-47}$ | 3.198x10$^{-26}$ | 2.167x10$^{25}$ | 2.386x10$^{-10}$ |
| $^{326}$124 | $^{24}$Ne | $^{302}$114 | 83.260 | 5.387x10$^{-53}$ | 1.215x10$^{-31}$ | 5.704x10$^{30}$ | 7.000x10$^{-15}$ |
| $^{328}$124 | $^{24}$Ne | $^{304}$114 | 82.040 | 5.877x10$^{-55}$ | 1.306x10$^{-33}$ | 5.306x10$^{32}$ | 1.108x10$^{-19}$ |
| $^{330}$124 | $^{24}$Ne | $^{306}$114 | 81.580 | 1.286x10$^{-55}$ | 2.842x10$^{-34}$ | 2.438x10$^{33}$ | 9.686x10$^{-25}$ |
| $^{332}$124 | $^{24}$Ne | $^{308}$114 | 81.660 | 2.458x10$^{-55}$ | 5.438x10$^{-34}$ | 1.274x10$^{33}$ | 4.796x10$^{-30}$ |
| $^{334}$124 | $^{24}$Ne | $^{310}$114 | 81.600 | 2.647x10$^{-55}$ | 5.851x10$^{-34}$ | 1.184x10$^{33}$ | 1.375x10$^{-35}$ |

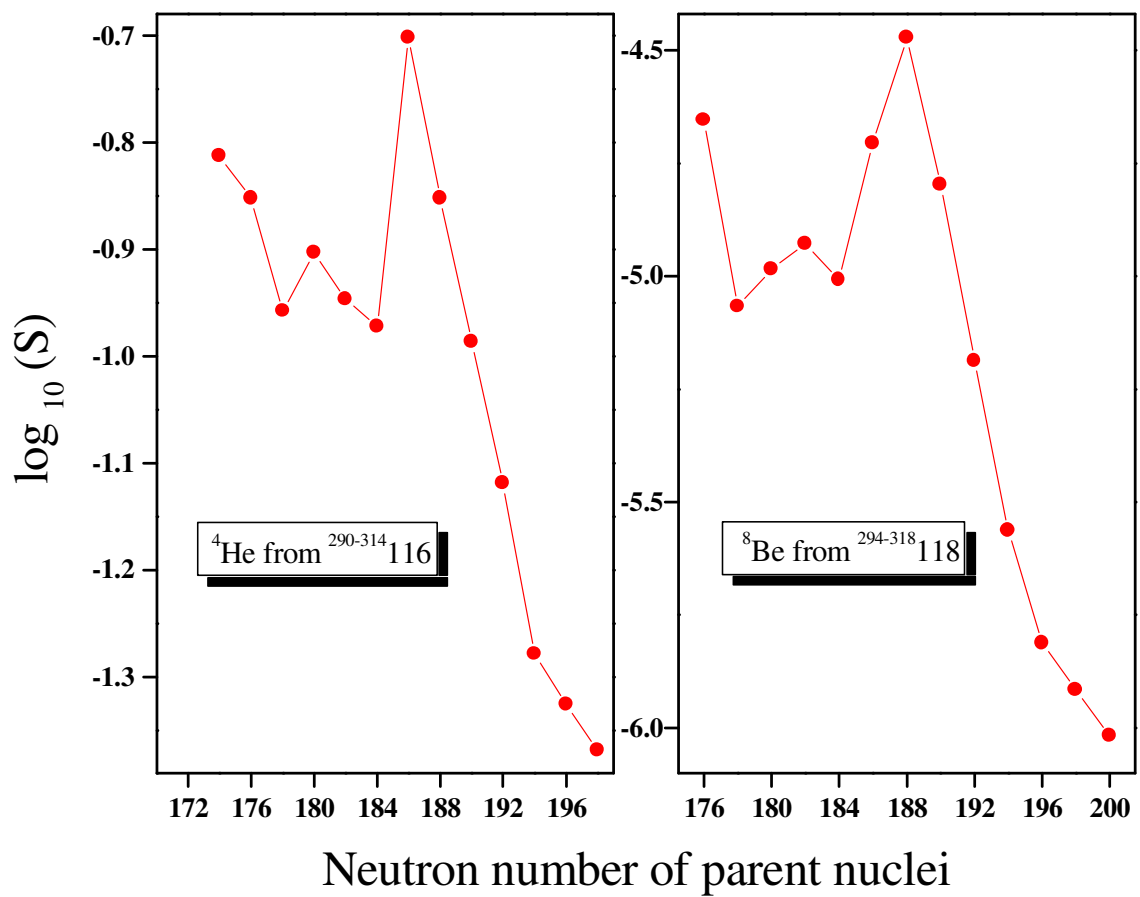

**Figure 1.** The computed $\log_{10}(S)$ values plotted against the neutron number of the parent, for the emission of clusters $^4$He and $^8$Be from $^{290-314}$116 and $^{294-318}$118 isotopes respectively.

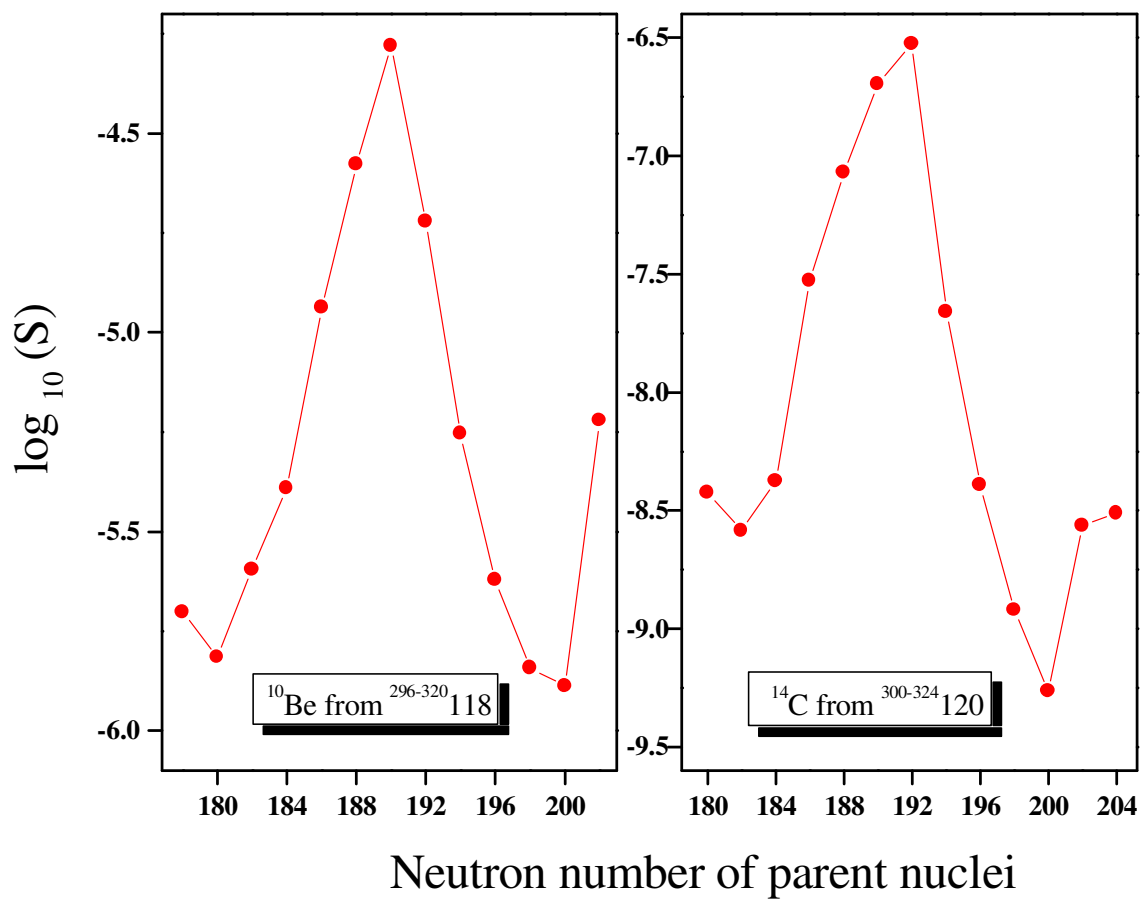

**Figure 2.** The computed $\log_{10}(S)$ values plotted against the neutron number of the parent, for the emission of clusters $^{10}$Be and $^{14}$C from $^{296-320}$118 and $^{300-324}$120 isotopes respectively.

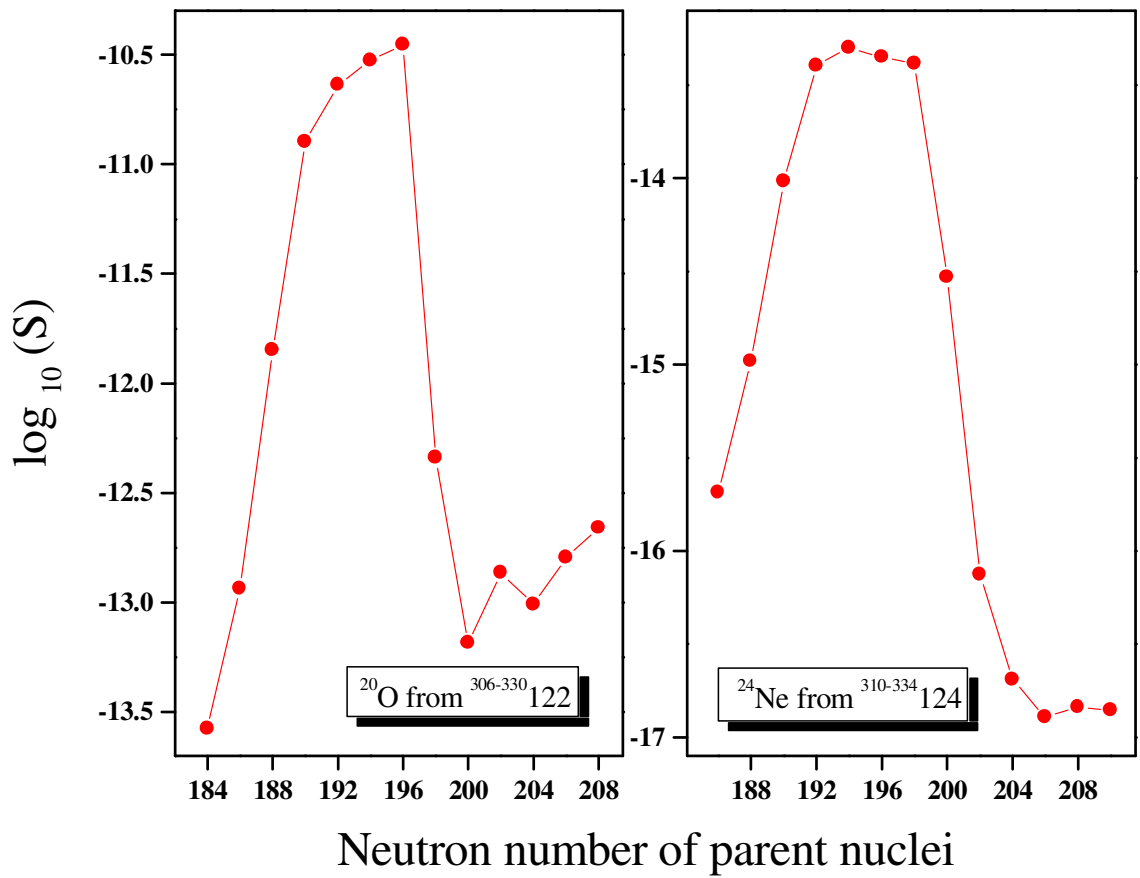

**Figure 3.** The computed $\log_{10}(S)$ values plotted against the neutron number of the parent, for the emission of clusters $^{20}$O and $^{24}$Ne from $^{306-330}$120 and $^{310-334}$124 isotopes respectively.

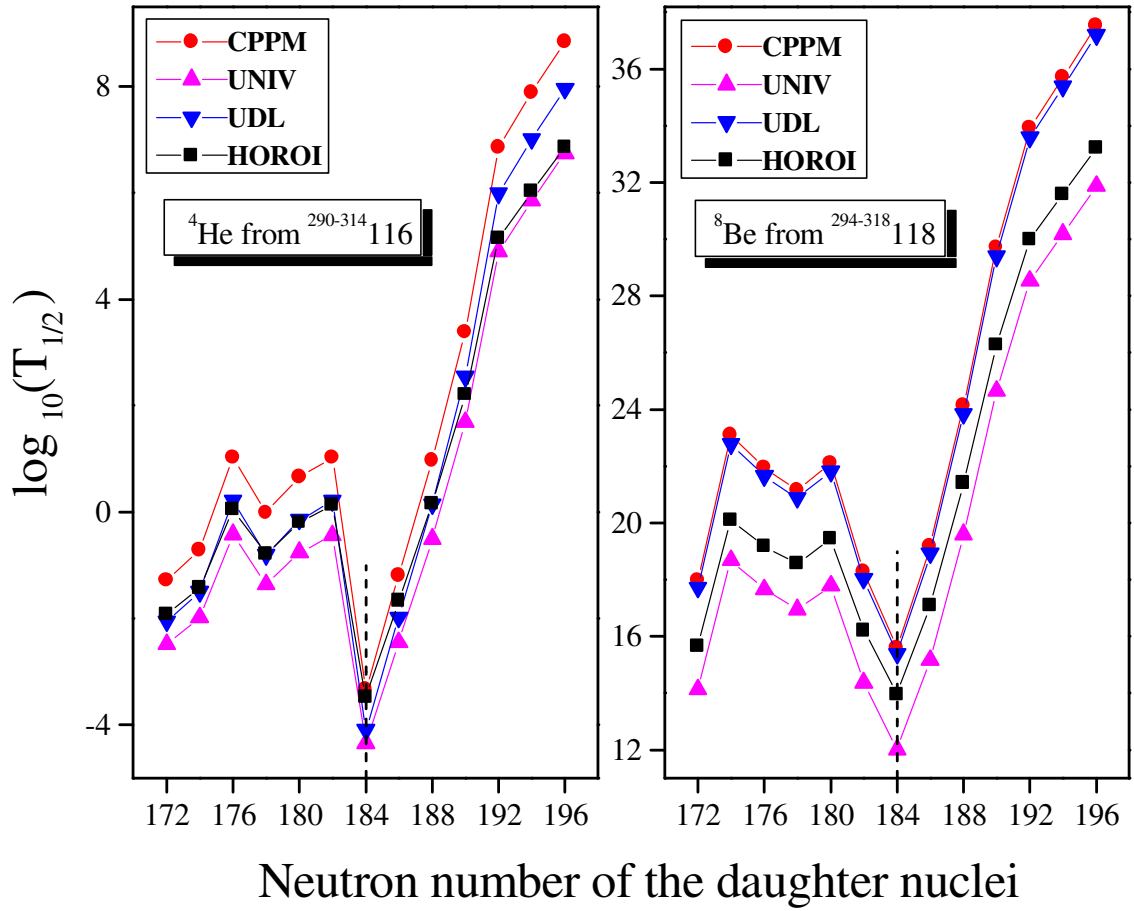

**Figure 4.** The computed $\log_{10}(T_{1/2})$ values vs. neutron number of daughter for the emission of clusters $^4$He and $^8$Be from $^{290-314}$116 and $^{294-318}$118 isotopes respectively. $T_{1/2}$ is in seconds.

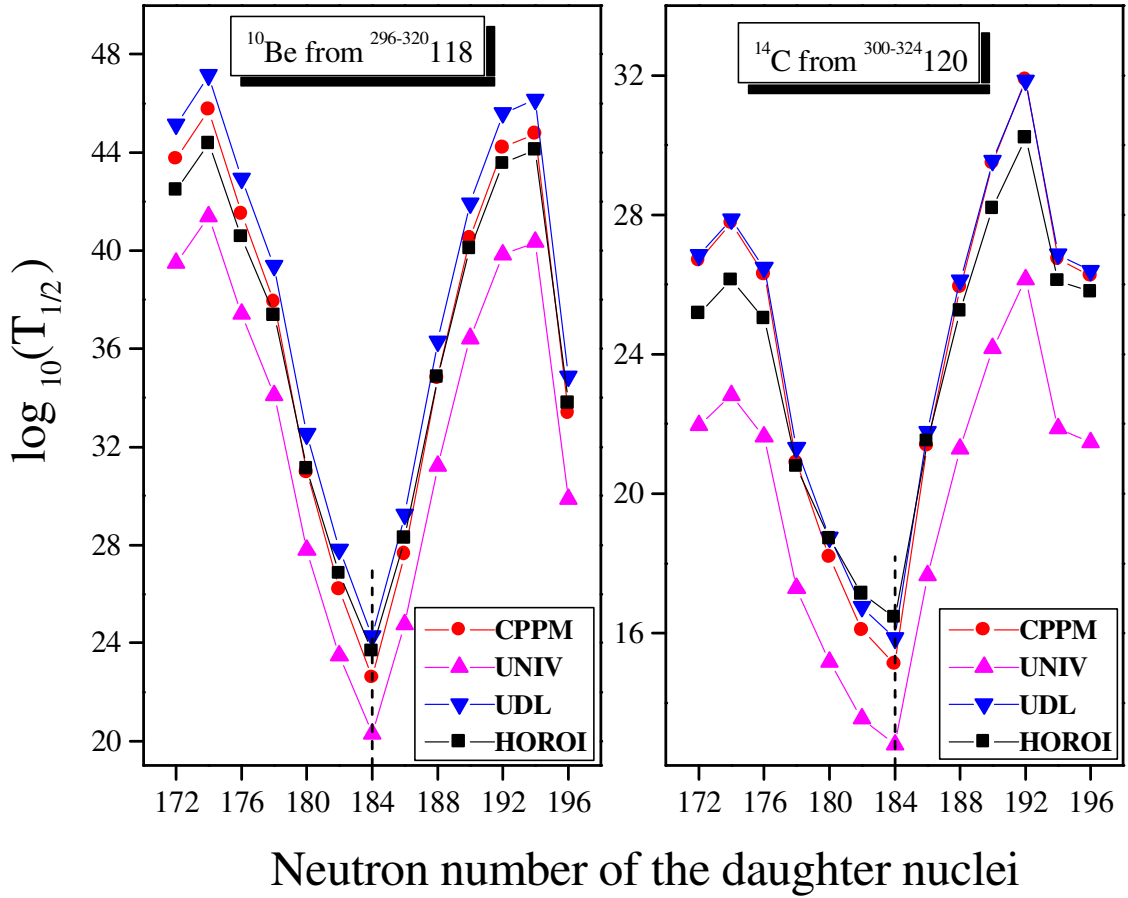

**Figure 5.** The computed $\log_{10}(T_{1/2})$ values vs. neutron number of daughter for the emission of clusters $^{10}$Be and $^{14}$C from $^{296-320}$118 and $^{300-344}$120 isotopes respectively. $T_{1/2}$ is in seconds.

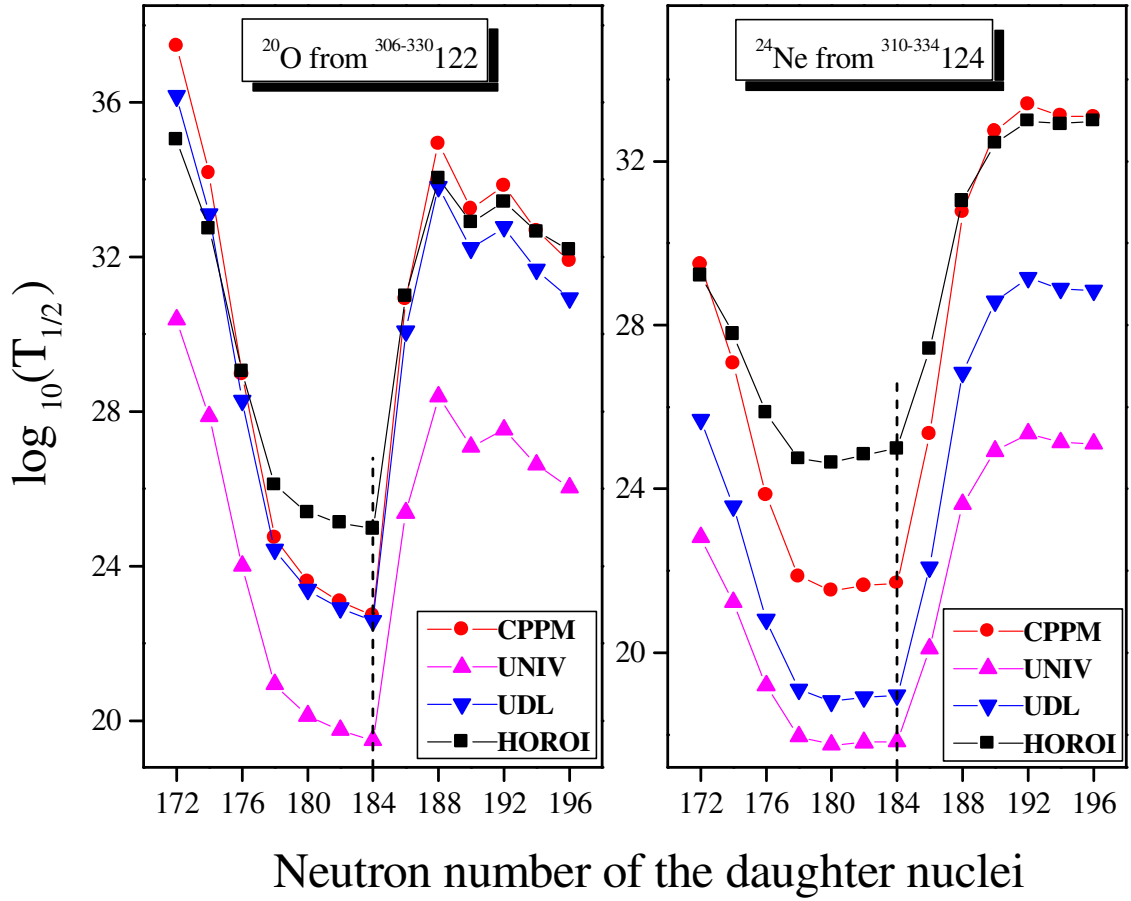

**Figure 6.** The computed $\log_{10}(T_{1/2})$ values vs. neutron number of daughter for the emission of clusters $^{20}$O and $^{24}$Ne from $^{306-330}$120 and $^{310-334}$124 isotopes respectively. $T_{1/2}$ is in seconds.